\documentclass[12pt,doc,english]{apa6}
\usepackage[T1]{fontenc}
\usepackage[latin9]{inputenc}
\usepackage{listings}

\usepackage{verbatim}
\usepackage{bm}
\usepackage{amsmath}
\usepackage{amssymb}
\usepackage{graphicx}
\usepackage[authoryear]{natbib}

\makeatletter

\providecommand{\tabularnewline}{\\}
\newcommand{\lyxdot}{.}

\usepackage{threeparttable}
\usepackage{multirow}
\usepackage{rotating}
\usepackage{subfig}
\usepackage{bm}
\usepackage{rotating}
\usepackage{arydshln}
\usepackage{times}

\@ifundefined{showcaptionsetup}{}{%
 \PassOptionsToPackage{caption=false}{subfig}}
\usepackage{subfig}
\makeatother

\usepackage{babel}
\begin{document}

\title{Mediation analysis with missing data through multiple imputation
and bootstrap}

\author{Lijuan Wang, Zhiyong Zhang, and Xin Tong}

\affiliation{University of Notre Dame}
\authornote{Correspondance should be sent to Zhiyong Zhang, 118 Haggar Hall, Univeristy of Notre Dame, IN 46556. Email: zzhang4@nd.edu.}

\shorttitle{mediation analysis with missing data}

\abstract{A method using multiple imputation and bootstrap for dealing with
missing data in mediation analysis is introduced and implemented in
SAS. Through simulation studies, it is shown that the method performs
well for both MCAR and MAR data without and with auxiliary variables.
It is also shown that the method works equally well for MNAR data
if auxiliary variables related to missingness are included. The application
of the method is demonstrated through the analysis of a subset of
data from the National Longitudinal Survey of Youth.}

\keywords{Mediation analysis, missing data, multiple imputation, auxiliary
variables, bootstrap, SAS}

\maketitle

\section{Introduction}

Mediation models and mediation analysis are widely used in behavioral
and social sciences as well as in health and medical research. The
influential article on mediation analysis by \citet{baron1986} has
been cited more than 8,000 times. Mediation models are very useful
for theory development and testing as well as for identification of
intervention points in applied work. Although mediation models were
first developed in psychology \citep[e.g., ][]{MacCorquodale1948,Woodworth1928},
they have been recognized and used in many disciplines where the mediation
effect is also known as the indirect effect \citep[Sociology, ][]{Alwin75a}
and the surrogate or intermediate endpoint effect \citep[Epidemiology, ][]{Freeman1992}. 

Figure \ref{fig:mediation} \citep[after ][]{Shrout2002} depicts
the path diagram of a simple mediation model. In this figure, $X$,
$M$, and $Y$ represent the independent or input variable, the mediation
variable (mediator), and the dependent or outcome variable, respectively.
The $e_{M}$ and $e_{Y}$ are residuals or disturbances with variances
$\sigma_{eM}^{2}$ and $\sigma_{eY}^{2}$. $c'$ is called the direct
effect and the mediation effect or indirect effect is measured by
the product term $ab$. The other parameters in this model include
the intercepts $i_{M}$ and $i_{Y}$. 

\begin{center}
\begin{figure}[h]
\begin{centering}
\includegraphics[scale=0.7]{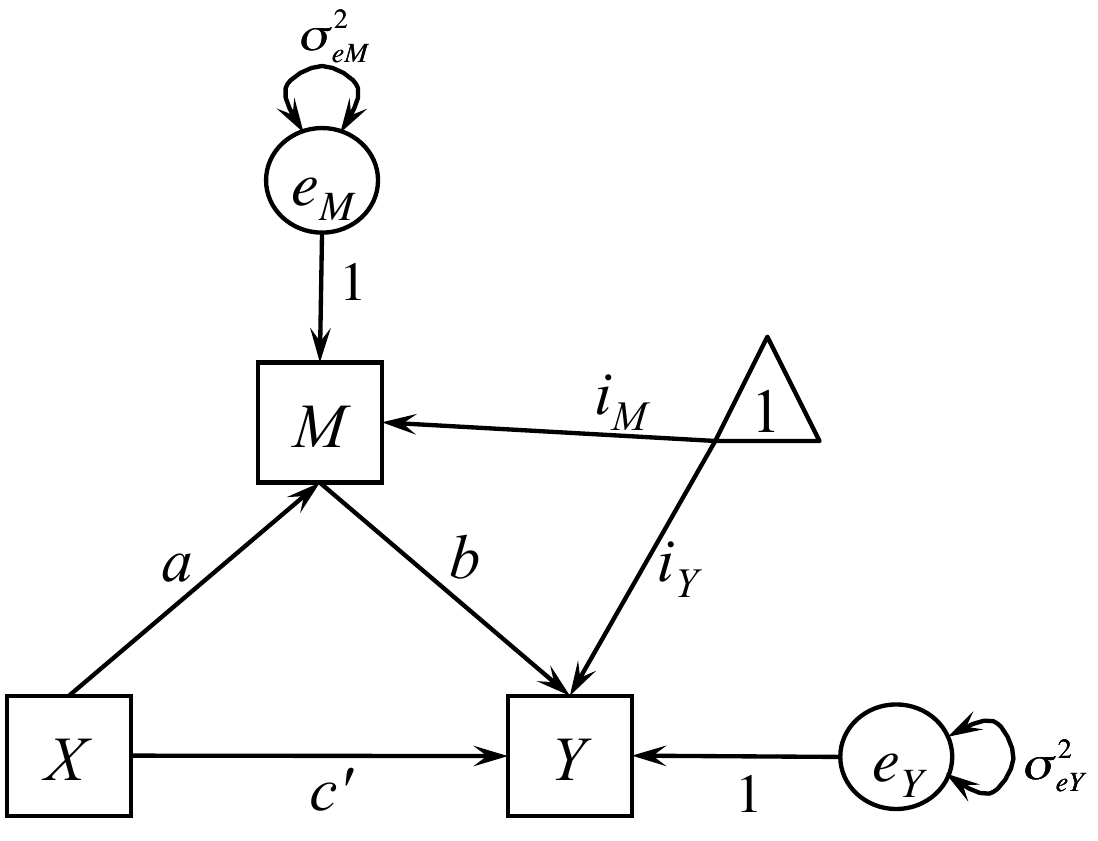}
\par\end{centering}

\caption{Path diagram demonstration of a mediation model.}
\label{fig:mediation}
\end{figure}

\par\end{center}

Statistical approaches to estimating and testing mediation effects
with complete data have been discussed extensively in the psychological
literature \citep[e.g., ][]{baron1986,bollen1990,mackinnon2002,MacKinnon2007,Shrout2002}.
One way to test mediation effects is to test $H_{0}:ab=0$. If a large
sample is available, the normal approximation method can be used,
which constructs the standard error of $ab$ through the delta method
so that $s.e.(ab)=\sqrt{\hat{b}^{2}\hat{\sigma_{a}^{2}}+2\hat{a}\hat{b}\hat{\sigma}_{ab}+\hat{a}^{2}\hat{\sigma_{b}^{2}}}$
with parameter estimates $\hat{a}$ and $\hat{b}$, their estimated
variances $\hat{\sigma_{a}^{2}}$ and $\hat{\sigma_{b}^{2}}$, and
covariance $\hat{\sigma}_{ab}$ \citep[e.g., ][]{sobel1982,sobel1986}.
Many researchers suggested that the distribution of $ab$ may not
be normal especially when the sample size is small although with large
sample sizes the distribution may approach normality \citep{bollen1990,mackinnon2002}.
Thus, bootstrap methods have been recommended to obtain the empirical
distribution and confidence interval of $ab$ \citep{MacKinnon2004,mallinckrodt2006,Preacher2008,Shrout2002,zhangwang2007a}. 

Mediation analysis can be conducted in a variety of programs and software.
Notably, the SAS and SPSS macros by \citet{Preacher2004,Preacher2008}
have popularized the application of bootstrap techniques in mediation
analysis. Based on search results from Google scholar, \citet{Preacher2004}
has been cited more than 900 times and \citet{Preacher2008} has already
been cited more than 400 times in less than two years after publication.

Missing data problem is continuously a challenge even for a well designed
study. Although there are approaches to dealing with missing data
for path analysis in general \citep[for a recent review, see ][]{Graham2009},
there are few studies focusing on the treatment of missing data in
mediation analysis. Particularly, mediation analysis is different
from typical path analysis because the focus is on the product of
two path coefficients. A common practice is to analyze complete data
through listwise deletion or pairwise deletion \citep[e.g., ][]{Chen2005457,Preacher2004}.
However, with the availability of advanced approaches such as multiple
imputation (MI), listwise and pairwise deletion is no longer deemed
acceptable \citep{Little2002,Savalei2009,Schafer1997}.

In this study, we discuss how to deal with missing data for mediation
analysis through multiple imputation (MI) and bootstrap using SAS.
The rationale of using multiple imputation is that it can be implemented
in existing popular statistical software such as SAS and it can deal
with different types of missing data. In the following, we will first
present the technical backgrounds of multiple imputation for mediation
analysis with missing data. Then, we will discuss how to implement
the method in SAS. After that, we will present several simulation
examples to evaluate the performance of MI for mediation analysis
with missing data. Finally, an empirical example will be used to demonstrate
the application of the method.

\section{Method}

In this section, we present the technical backgrounds of mediation
analysis with missing data through multiple imputation and bootstrap.
First, we will discuss how to estimate mediation model parameters
with complete data. Second, we will reiterate the definition of missing
data mechanisms by \citet{Little2002}. Third, we will discuss how
to apply multiple imputation to mediation analysis. Finally, we will
discuss the bootstrap procedure to obtain the bias corrected confident
intervals for mediation model parameters.

\subsection{Complete data mediation analysis}

In mathematical form, the mediation model displayed in Figure \ref{fig:mediation}
can be expressed using two equations,
\begin{align}
M & =i_{M}+aX+e_{M}\nonumber \\
Y & =i_{Y}+bM+c'X+e_{Y},\label{eq:mediation}
\end{align}
which can be viewed as a collection of two linear regression models.
To obtain the parameter estimates in the model, one can maximize the
product of the likelihood functions from the two regression models
using the maximum likelihood method. Because $e_{M}$ and $e_{Y}$
are assumed to be independent, maximizing the product of the likelihood
functions is equivalent to maximizing the likelihood function of each
regression model separately. Thus, parameter estimates can be obtained
by fitting two separate regression models in Equation \ref{eq:mediation}.
Specifically, the mediation effect estimate is $\hat{a}\hat{b}$ with
\begin{align}
\hat{a} & =s_{XM}/s_{X}^{2}\nonumber \\
\hat{b} & =(s_{MY}s_{X}^{2}-s_{XM}s_{XY})/(s_{X}^{2}s_{M}^{2}-s_{XM}^{2})\label{eq:estimation}
\end{align}
where $s_{X}^{2},s_{M}^{2},s_{Y}^{2},s_{XM},s_{MY},s_{XY}$ are sample
variances and covariances of $X,M,Y$, respectively.

\subsection{Missing mechanisms}

\citet{Little87a,Little2002} have distinguished three types of missing
data -- missing completely at random (MCAR), missing at random (MAR),
and missing not at random (MNAR). Let $D=(X,M,Y)$ denote all data
that can be potentially observed in a mediation model. $D_{obs}$
and $D_{miss}$ denote data that are actually observed and data that
are not observed, respectively. Let $R$ denote an indicator matrix
of zeros and ones. If a datum in $D$ is missing, the corresponding
element in $R$ is equal to 1. Otherwise, it is equal to 0. Finally,
let $A$ denote the auxiliary variables that are related to the missingness
of $D$ but not a component of the mediation model in Equation \ref{eq:mediation}.

If the missing mechanism is MCAR, then we have 
\[
\Pr(R|D_{obs},D_{miss},A,\bm{\theta})=\Pr(R|\bm{\theta}),
\]
where the vector $\bm{\theta}$ represents all model parameters in
the mediation model including $a$, $b$, $ab$, $c'$, $i_{M}$,
$i_{Y}$, $\sigma_{eM}^{2}$, and $\sigma_{eY}^{2}$. This suggests
that missing data $D_{miss}$ are a simple random sample of $D$ and
missingness is not related to the data of interest $D$ or auxiliary
variables A. 

If the missing mechanism is MAR, then 
\[
\Pr(R|D_{obs},D_{miss},A,\bm{\theta})=\Pr(R|D_{obs},\bm{\theta}),
\]
which indicates that the probability that a datum is missing is related
to the observed data $D_{obs}$ but not to the missing data $D_{miss}$. 

Finally, if the probability that a datum is missing is related to
the missing data $D_{miss}$ or auxiliary variables $A$ while $A$
are not considered in the data analysis, the missing mechanism is
MNAR.

\subsection{Multiple imputation for mediation analysis with missing data}

Most techniques dealing with missing data including multiple imputation
in general require missing data to be either MCAR or MAR \citep[see also, e.g., ][]{Little2002,Schafer1997}.
For MNAR, the missing mechanism has to be known to correctly recover
model parameters. Practically, researchers have suggested including
auxiliary variables to facilitate MNAR missing data analysis \citep{Graham2003,Savalei2009}.
Auxiliary variables are variables that are not a component of a model
(not model variables) but can explain missingness of variables in
the model. After including appropriate auxiliary variables, we may
be able to assume that data from both model variables and auxiliary
variables are MAR.

The setting for mediation analysis with missing data is described
below. Assume that a set of $p(p\geq0)$ auxiliary variables $A_{1},A_{2},\ldots,A_{p}$
are available. These auxiliary variables may or may not be related
to missingness of the mediation model variables. Furthermore, there
may or may not be missing data in auxiliary variables. By augmenting
the auxiliary variables with the mediation model variables, we have
a total of $p+3$ variables denoted by $D=(X,M,Y,A_{1},\ldots,A_{p})$.
To proceed, we assume that the missing mechanism is MAR after including
the auxiliary variables. That is 
\[
\Pr(R|D_{obs},D_{miss},A_{1},\ldots,A_{p},\bm{\theta})=\Pr(R|D_{obs},A_{1},\ldots,A_{p},\bm{\theta}).
\]

Multiple imputation \citep{Little2002,Rubin76a,Schafer1997} is a
procedure to fill each missing value with a set of plausible values.
The multiple imputed data sets are then analyzed using standard procedures
for complete data and the results from these analyses are combined
for obtaining point estimates of model parameters and standard errors
of parameter estimates. For mediation analysis with missing data,
the following steps can be implemented for obtaining point estimates
of mediation model parameters.
\begin{APAenumerate}
\item Assuming that $D=(X,M,Y,A_{1},\ldots,A_{p})$ are from a multivariate
normal distribution, generate $K$ ($K$ is the number of multiple
imputations) sets of values for each missing value. Combine the generated
values with the observed data to produce $K$ sets of complete data
\citep{Schafer1997}. 
\item For each of the $K$ sets of complete data, apply the formula in Equation
\ref{eq:estimation} to obtain a point mediation effect estimate $\hat{a}_{k}\hat{b}_{k}(j=1,\ldots,K)$. 
\item The point estimate for the mediation effect through multiple imputation
is the average of the $K$ complete data mediation effect estimates:
\[
\hat{a}\hat{b}=\frac{1}{K}\sum_{k=1}^{K}\hat{a}_{k}\hat{b}_{k}.
\]

\end{APAenumerate}
Parameter estimates for the other model parameters $a$, $b$, $c'$,
$i_{M}$, $i_{Y}$, $\sigma_{eM}^{2}$, and $\sigma_{eY}^{2}$ can
be obtained in the same way.

\subsection{Testing mediation effects through the bootstrap method }

The procedure described above is implemented to obtain point estimates
of mediation effects. To test mediation effects, we need to obtain
standard errors of the parameter estimates. Because mediation effects
are measured by $ab$, researchers suggest using bootstrap to obtain
empirical standard errors as mentioned in a previous section. The
bootstrap method \citep{efron1979,efron1987} was first employed in
mediation analysis by \citet{bollen1990} and has been studied in
a variety of research contexts \citep[e.g., ][]{MacKinnon2004,mallinckrodt2006,Preacher2008,Shrout2002}.
This method has no distribution assumption on the indirect effect
$ab$. Instead, it approximates the distribution of $ab$ using its
bootstrap empirical distribution.

The bootstrap method used in \citet{bollen1990} can be applied along
with multiple imputation to obtain standard errors of mediation effect
estimates and confidence intervals for mediation analysis with missing
data. Specifically, the following procedure can be used. 
\begin{APAenumerate}
\item Using the \emph{original data set} (Sample size = \emph{N}) as a population,
draw a bootstrap sample of \emph{N} persons randomly with replacement
from the original data set. This bootstrap sample generally would
contain missing data.
\item With the bootstrap sample, implement the $K$ multiple imputation
procedure described in the above section to obtain point estimates
of model parameters and a point estimate of the mediation effect . 
\item Repeat Steps 1 and 2 for a total of $B$ times. $B$ is called the
number of bootstrap samples. 
\item Empirical distributions of model parameters and the mediation effect
are then obtained using the $B$ sets of bootstrap point estimates.
Thus, confidence intervals of model parameters and mediation effect
can be constructed. 
\end{APAenumerate}
The procedure described above can be considered as a procedure of
$K$ multiple imputations nested within $B$ bootstrap samples. Using
the $B$ bootstrap sample point estimates, one can obtain bootstrap
standard errors and confidence intervals of model parameters and mediation
effects conveniently. Let $\bm{\theta}=(iM,iY,a,b,c',\sigma_{eM}^{2},\sigma_{eY}^{2},ab)^{t}$
denote a vector of model parameters and the mediation effect $ab$.
With data from each bootstrap, we can obtain $\hat{\bm{\theta}}^{b},\ b=1,\ldots,B$.
The standard error of the $p$th parameter $\hat{{\theta}}_{p}$ can
be calculated as 
\[
\widehat{s.e.(\hat{\theta}_{p})}=\sqrt{\sum_{b=1}^{B}(\hat{\theta}_{p}^{b}-\bar{\hat{\theta}}_{p}^{b})^{2}/(B-1)}
\]
 with 
\[
\bar{\hat{\theta}}_{p}^{b}=\sum_{b=1}^{B}\hat{\theta}_{p}^{b}/B.
\]

Many methods for constructing confidence intervals from $\hat{\bm{\theta}}^{b}$
have been proposed such as the percentile interval, the bias-corrected
(BC) interval, and the bias-corrected and accelerated (BCa) interval
\citep{efron1987,MacKinnon2004}. In the present study, we focus on
the BC interval because \citet{MacKinnon2004} showed that the BC
confidence intervals have correct Type I error and largest power among
many different evaluated confidence intervals.

The $1-2\alpha$ BC interval for the $p$th element of $\bm{\theta}$
can be constructed using the percentiles $\hat{\theta}_{p}^{b}(\tilde{\alpha}_{l})$
and $\hat{\theta}_{p}^{b}(\tilde{\alpha}_{u})$ of $\hat{\theta}_{p}^{b}$.
Here 
\[
\tilde{\alpha}_{l}=\Phi(2z_{0}+z^{(\alpha)})
\]
 and 
\[
\tilde{\alpha}_{u}=\Phi(2z_{0}+z^{(1-\alpha)})
\]
 where $\Phi$ is the standard cumulative normal distribution function
and $z^{(\alpha)}$ is the $\alpha$ percentile of the standard normal
distribution and 
\[
z_{0}=\Phi^{-1}\left[\frac{\text{number of times that }\hat{\theta}_{p}^{b}<\hat{\theta}_{p}}{B}\right].
\]

\section{Multiple imputation and bootstrap for mediation analysis with missing
data in SAS}

To facilitate the implementation of the method described in the above
section, we have written a SAS program for mediation analysis with
missing data using multiple imputation and bootstrap. The complete
SAS program scripts are contained in the Appendix. Now we briefly
explain the functioning of each part of the SAS program. 

Lines 3-9 of the SAS program specifies all global parameters that
control multiple imputation and bootstrap for mediation analysis.
This part is the one that a user needs to modify according to his/her
data analysis environment. Line 3 specifies the directory and name
of the data file to be used. Line 4 lists the names of the variables
in the data file. Line 5 specifies the missing data value indicator.
For example, 99999 in the data file represents a missing datum. Line
6 specifies the number of imputations ($K$) for imputing missing
data. Line 7 defines the number of bootstrap samples ($B$). A number
larger than 1000 is usually recommended. Line 8 and Line 9 specify
the confidence level and the random number generator seed, respectively. 

Lines 15-22 first read data into SAS from the data file specified
on line 3 and then change missing data to the SAS missing data format
- a dot. Lines 26-28 impute missing data for the original data set
with auxiliary variables and generate $K$ imputed data sets. Lines
30-34 estimate the mediation model parameters for each imputed data
set. Lines 37-74 collect the results from the multiple imputed data
sets and save the point estimates of model parameters and mediation
effect in a SAS data set called ``\texttt{pointest}''. %
The SAS codes in this section produce point parameter estimates for
the model parameters and the mediation effect based on the original
data after multiple imputation.

Lines 77-88 generate $B$ bootstrap samples from the original data
set with the same sample size. Lines 91-95 impute each bootstrap sample
independently for $K$ times. Lines 98-143 produce point estimates
of mediation model parameters and mediation effect for each bootstrap
sample and collect the point estimates for all bootstrap samples in
the SAS data set named ``\texttt{bootest}''. 

The last part of the SAS program from Line 146 to Line 195 calculates
the bootstrap standard errors and the bias-corrected confidence intervals
for mediation model parameters and mediation effect. It also generates a
table containing the point estimates, standard errors, and confidence
intervals in the SAS output window.

To use the SAS program, one only needs to first change the global
parameters in Lines 3-9, usually only lines 3 and 4, and then run
the whole SAS program from the beginning to the end.

\section{Evaluating the method for mediation analysis with missing data}

In this section, we conduct several simulation studies to evaluate
the performance of the proposed method for mediation analysis with
missing data. We first evaluate its performance under different missing
data mechanisms including MCAR, MAR, and MNAR without and with auxiliary
variables. Then, we investigate how many imputations are needed for
mediation analysis with different proportions of missing data. In
the following, we first discuss our simulation design and then present
the simulation results.

\subsection{Simulation design}

For mediation analysis with complete data, simulation studies have
been conducted to investigate a variety of features of mediation models
\citep[e.g., ][]{mackinnon2002,MacKinnon2004}. For the current study,
we follow the parameter setup from previous literature and set the
model parameter values to be $a=b=.39$, $c'=0$, $i_{M}=i_{Y}=0$,
and $\sigma_{eM}^{2}=\sigma_{eY}^{2}=\sigma_{eX}^{2}=1$. Furthermore
, we fix the sample size at $N=100$ and consider three proportions
of missingness with missing data percentages at 10\%, 20\%, and 40\%,
respectively. To facilitate the comparisons among different missing
mechanisms, missing data are only allowed in $M$ and $Y$ although
our SAS program allows missingness in $X$. Two auxiliary variables
($A_{1}$ and $A_{2}$) are also generated where the correlation between
$A_{1}$ and $M$ and the correlation between $A_{2}$ and $Y$ are
both $0.5$. For each of the following simulation studies, results
are from $R=1,000$ sets of simulated data.%

For each simulation study, we report point estimate bias, coverage
probability, and power or Type I error for evaluations. Let $\theta$
denote the true parameter value in the simulation and $\hat{\theta}_{r}(r=1,\ldots,1000)$
denote the corresponding estimate from the $r$th replication. The
bias is calculated as 
\[
\text{Bias}=\begin{cases}
100\times\left[\frac{\sum_{r=1}^{1000}\hat{\theta}_{r}}{1000\theta}-1\right] & \theta\neq0\\
100\times\left[\frac{\sum_{r=1}^{1000}\hat{\theta}_{r}}{1000}-\theta\right] & \theta=0
\end{cases}.
\]
Note that the bias is rescaled by multiplying 100. Smaller bias indicates
the point estimate is less biased. Furthermore, Let $\hat{l_{r}}$
and $\hat{u_{r}}$ denote the lower and upper limits of the $95\%$
confidence interval in the $r$th replication. The coverage probability
is calculated by 
\[
\text{coverage}=\frac{\#(\hat{l_{r}}<\gamma<\hat{u}_{r})}{1000}
\]
where $\#(\hat{l_{r}}<\gamma<\hat{u}_{r})$ is the total number of
replications with confidence intervals covering the true parameter
value. Good $95\%$ confidence intervals should give coverage probabilities
close to $0.95$. Power or Type I error is calculated by
\[
\text{power}=\frac{\#(\hat{l_{r}}>0)+\#(\hat{u}_{r}<0)}{1000}
\]
where $\#(\hat{l_{i}}>0)$ is the total number of replications with
the lower limits of confidence intervals larger than 0 and $\#(\hat{u}_{r}<0)$
is the total number of replications with the upper limits smaller
than 0. If the population parameter value is not equal to 0, a better
method should have greater statistical power. If the population parameter
value is equal to 0, a good method should have type I error close
to the nominal alpha level.

\subsection{Simulation 1. Analysis of MCAR data}

The parameter estimate biases, coverage probabilities, and power/Type
I errors for MCAR data with 10\%, 20\%, and 40\% missing data are
obtained without and with auxiliary variables and are summarized in
Table \ref{tab:mcar}. From the results, we can conclude the following.
First, biases of the parameter estimates for all conditions under
the studied MCAR conditions are smaller than 1.5\%. Second, the coverage
probabilities are close to the true value .95 except that the coverage
probabilities of variance parameters range from .88 to .94 and are
slightly underestimated. Third, the inclusion of auxiliary variables
in MCAR data mediation analysis does not seem to influence the accuracy
of parameter estimates and coverage probabilities although the auxiliary
variables are correlated with $M$ and $Y$($r=.5$). The use of auxiliary
variables, however, slightly boosters the power of detecting mediation
effect especially when the missing proportion is larger (e.g., 40\%).

\begin{table}[h]
\caption{Biases, coverage probabilities, and power/type I error under MCAR
situations}

\label{tab:mcar} 

\begin{tabular}{c|c|ccc|ccc}
\hline 
 &  & \multicolumn{3}{c|}{Without Auxiliary Variables} & \multicolumn{3}{c}{With Auxiliary Variables }\tabularnewline
\hline 
 &  & Bias & Coverage & Power/Type I error & Bias & Coverage & Power/Type I error\tabularnewline
\hline 
\multirow{8}{*}{10\%} & $a$ & 0.595 & 0.938 & 0.946 & 0.861 & 0.943 & 0.956\tabularnewline
 & $b$ & 0.055 & 0.941 & 0.920 & -0.130 & 0.944 & 0.927\tabularnewline
 & $c'$ & 0.487 & 0.953 & 0.047 & 0.304 & 0.945 & 0.055\tabularnewline
 & $ab$ & 0.219 & 0.967 & 0.900 & 0.263 & 0.967 & 0.920\tabularnewline
 & $i_{Y}$ & 0.116 & 0.945 & 0.055 & 0.304 & 0.946 & 0.054\tabularnewline
 & $i_{M}$ & 0.065 & 0.956 & 0.044 & -0.072 & 0.952 & 0.048\tabularnewline
 & $\sigma_{eY}^{2}$ & -0.657 & 0.935 & 1.000 & -0.494 & 0.931 & 1.000\tabularnewline
 & $\sigma_{eM}^{2}$ & -0.148 & 0.931 & 1.000 & -0.051 & 0.930 & 1.000\tabularnewline
\hline 
\multirow{8}{*}{20\%} & $a$ & -0.218 & 0.936 & 0.907 & -0.047 & 0.938 & 0.920\tabularnewline
 & $b$ & -0.525 & 0.940 & 0.829 & -0.131 & 0.937 & 0.862\tabularnewline
 & $c'$ & 0.430 & 0.934 & 0.066 & 0.266 & 0.941 & 0.059\tabularnewline
 & $ab$ & -1.222 & 0.966 & 0.808 & -0.593 & 0.963 & 0.845\tabularnewline
 & $i_{Y}$ & -0.165 & 0.946 & 0.054 & -0.204 & 0.944 & 0.056\tabularnewline
 & $i_{M}$ & 0.349 & 0.956 & 0.044 & 0.268 & 0.954 & 0.046\tabularnewline
 & $\sigma_{eY}^{2}$ & -0.818 & 0.920 & 1.000 & -0.539 & 0.918 & 1.000\tabularnewline
 & $\sigma_{eM}^{2}$ & -0.244 & 0.942 & 1.000 & -0.105 & 0.944 & 1.000\tabularnewline
\hline 
\multirow{8}{*}{40\%} & $a$ & 0.640 & 0.938 & 0.822 & 0.634 & 0.930 & 0.849\tabularnewline
 & $b$ & -1.310 & 0.930 & 0.565 & -0.593 & 0.935 & 0.635\tabularnewline
 & $c'$ & 0.607 & 0.944 & 0.056 & 0.226 & 0.945 & 0.055\tabularnewline
 & $ab$ & -0.716 & 0.946 & 0.531 & 0.112 & 0.950 & 0.615\tabularnewline
 & $i_{Y}$ & -0.007 & 0.943 & 0.057 & -0.044 & 0.939 & 0.061\tabularnewline
 & $i_{M}$ & -0.127 & 0.966 & 0.034 & 0.050 & 0.963 & 0.037\tabularnewline
 & $\sigma_{eY}^{2}$ & -1.484 & 0.888 & 1.000 & -0.860 & 0.911 & 1.000\tabularnewline
 & $\sigma_{eM}^{2}$ & 0.077 & 0.924 & 1.000 & 0.498 & 0.933 & 1.000\tabularnewline
\hline 
\end{tabular}
\end{table}

\subsection{Simulation 2. Analysis of MAR data}

The estimate biases, coverage probabilities, and power for MAR data
analysis are summarized in Table \ref{tab:mar}. The findings from
MAR data are similar to those from MCAR data and thus are not repeated
here. However, the power of detecting mediation effects from MAR data
are smaller than that from MCAR data given the same proportion of
missing data.

\begin{table}[htbp]
\caption{Biases, coverage probabilities, and power/type I error under MAR situations}

\label{tab:mar}

\begin{tabular}{c|c|ccc|ccc}
\hline 
 &  & \multicolumn{3}{c|}{Without Auxiliary Variables} & \multicolumn{3}{c}{With Auxiliary Variables }\tabularnewline
\hline 
 &  & Bias & Coverage & Power/type I error & Bias & Coverage & Power/type I error\tabularnewline
\hline 
\multirow{8}{*}{10\%} & $a$ & 0.716 & 0.944 & 0.927 & 0.439 & 0.938 & 0.929\tabularnewline
 & $b$ & -0.331 & 0.936 & 0.896 & -0.314 & 0.946 & 0.917\tabularnewline
 & $c'$ & 0.679 & 0.954 & 0.046 & 0.435 & 0.947 & 0.053\tabularnewline
 & $ab$ & -0.119 & 0.957 & 0.870 & -0.403 & 0.961 & 0.893\tabularnewline
 & $i_{Y}$ & 0.369 & 0.948 & 0.052 & 0.294 & 0.948 & 0.052\tabularnewline
 & $i_{M}$ & -0.010 & 0.956 & 0.044 & -0.084 & 0.956 & 0.044\tabularnewline
 & $\sigma_{eY}^{2}$ & -0.574 & 0.924 & 1.000 & -0.457 & 0.921 & 1.000\tabularnewline
 & $\sigma_{eM}^{2}$ & -0.409 & 0.932 & 1.000 & -0.234 & 0.936 & 1.000\tabularnewline
\hline 
\multirow{8}{*}{20\%} & $a$ & 0.838 & 0.936 & 0.862 & -1.871 & 0.935 & 0.862\tabularnewline
 & $b$ & -0.897 & 0.935 & 0.807 & 0.095 & 0.933 & 0.833\tabularnewline
 & $c'$ & 0.320 & 0.940 & 0.060 & 0.027 & 0.952 & 0.048\tabularnewline
 & $ab$ & -0.546 & 0.962 & 0.767 & -1.940 & 0.958 & 0.791\tabularnewline
 & $i_{Y}$ & -0.294 & 0.945 & 0.055 & 0.035 & 0.952 & 0.048\tabularnewline
 & $i_{M}$ & 0.375 & 0.951 & 0.049 & -0.120 & 0.949 & 0.051\tabularnewline
 & $\sigma_{eY}^{2}$ & -0.886 & 0.918 & 1.000 & -0.650 & 0.920 & 1.000\tabularnewline
 & $\sigma_{eM}^{2}$ & -0.109 & 0.941 & 1.000 & -0.063 & 0.942 & 1.000\tabularnewline
\hline 
\multirow{8}{*}{40\%} & $a$ & -0.563 & 0.937 & 0.697 & -0.135 & 0.942 & 0.772\tabularnewline
 & $b$ & -1.863 & 0.929 & 0.597 & -1.372 & 0.926 & 0.647\tabularnewline
 & $c'$ & 0.837 & 0.940 & 0.060 & 0.315 & 0.945 & 0.055\tabularnewline
 & $ab$ & -2.932 & 0.960 & 0.511 & -1.747 & 0.955 & 0.599\tabularnewline
 & $i_{Y}$ & 0.220 & 0.950 & 0.050 & 0.004 & 0.943 & 0.057\tabularnewline
 & $i_{M}$ & -0.604 & 0.945 & 0.055 & -0.202 & 0.955 & 0.045\tabularnewline
 & $\sigma_{eY}^{2}$ & -1.137 & 0.908 & 1.000 & -0.452 & 0.909 & 1.000\tabularnewline
 & $\sigma_{eM}^{2}$ & -0.017 & 0.936 & 1.000 & 0.466 & 0.940 & 1.000\tabularnewline
\hline 
\end{tabular}
\end{table}

\subsection{Simulation 3. Analysis of MNAR data }

The results from MNAR data analysis are summarized in Table \ref{tab:mnar-1}.
The results clearly show that when auxiliary variables are not included,
parameter estimates are highly biased especially when the missing
data proportion is larger, e.g., about 67\% bias with 40\% missing
data for the mediation effect. Correspondingly, coverage probabilities
are highly underestimated. For example, with 40\% of missing data,
the coverage probabilities for intercepts and variance parameters
are almost zero. However, with the inclusion of appropriate auxiliary
variables, the parameter estimate biases dramatically decrease to
3\% or below and the coverage probabilities are close to 95\%. Thus,
multiple imputation can be used to analyze MNAR data and recover true
parameter values by including appropriate auxiliary variables that
can explain missingness of the variables in the mediation model.

\begin{table}
\caption{Biases, coverage probabilities, and power/type I error under MNAR
situations}

\label{tab:mnar-1}

\begin{tabular}{c|c|ccc|ccc}
\hline 
 &  & \multicolumn{3}{c|}{Without Auxiliary Variables} & \multicolumn{3}{c}{With Auxiliary Variables }\tabularnewline
\hline 
 &  & Bias & Coverage & Power/type I error & Bias & Coverage & Power/type I error\tabularnewline
\hline 
\multirow{8}{*}{10\%} & $a$ & -20.534 & 0.824 & 0.891 & 0.918 & 0.938 & 0.956\tabularnewline
 & $b$ & -15.339 & 0.888 & 0.827 & -1.099 & 0.933 & 0.923\tabularnewline
 & $c'$ & 1.930 & 0.955 & 0.045 & 0.468 & 0.946 & 0.054\tabularnewline
 & $ab$ & -32.633 & 0.831 & 0.800 & -0.513 & 0.951 & 0.925\tabularnewline
 & $i_{Y}$ & 11.320 & 0.739 & 0.261 & -0.076 & 0.951 & 0.049\tabularnewline
 & $i_{M}$ & 14.547 & 0.591 & 0.409 & -0.090 & 0.948 & 0.052\tabularnewline
 & $\sigma_{eY}^{2}$ & -13.029 & 0.532 & 1.000 & 0.004 & 0.939 & 1.000\tabularnewline
 & $\sigma_{eM}^{2}$ & -13.121 & 0.508 & 1.000 & 0.248 & 0.938 & 1.000\tabularnewline
\hline 
\multirow{8}{*}{20\%} & $a$ & -29.841 & 0.728 & 0.838 & 0.782 & 0.941 & 0.929\tabularnewline
 & $b$ & -27.443 & 0.810 & 0.589 & -2.856 & 0.928 & 0.826\tabularnewline
 & $c'$ & 2.197 & 0.943 & 0.057 & 0.190 & 0.947 & 0.053\tabularnewline
 & $ab$ & -49.117 & 0.673 & 0.570 & -2.583 & 0.941 & 0.815\tabularnewline
 & $i_{Y}$ & 22.228 & 0.356 & 0.644 & -0.001 & 0.955 & 0.045\tabularnewline
 & $i_{M}$ & 27.597 & 0.145 & 0.855 & 0.234 & 0.956 & 0.044\tabularnewline
 & $\sigma_{eY}^{2}$ & -20.661 & 0.228 & 1.000 & -0.494 & 0.933 & 1.000\tabularnewline
 & $\sigma_{eM}^{2}$ & -20.331 & 0.215 & 1.000 & 0.426 & 0.936 & 1.000\tabularnewline
\hline 
\multirow{8}{*}{40\%} & $a$ & -45.357 & 0.525 & 0.638 & -0.044 & 0.943 & 0.846\tabularnewline
 & $b$ & -38.421 & 0.839 & 0.355 & -1.824 & 0.934 & 0.666\tabularnewline
 & $c'$ & 3.041 & 0.936 & 0.064 & 1.053 & 0.947 & 0.053\tabularnewline
 & $ab$ & -66.815 & 0.559 & 0.305 & -2.951 & 0.951 & 0.642\tabularnewline
 & $i_{Y}$ & 45.112 & 0.113 & 0.887 & -1.212 & 0.950 & 0.050\tabularnewline
 & $i_{M}$ & 55.439 & 0.000 & 1.000 & -0.055 & 0.949 & 0.051\tabularnewline
 & $\sigma_{eY}^{2}$ & -31.444 & 0.086 & 1.000 & 0.333 & 0.923 & 1.000\tabularnewline
 & $\sigma_{eM}^{2}$ & -31.484 & 0.048 & 1.000 & 1.194 & 0.921 & 1.000\tabularnewline
\hline 
\end{tabular}
\end{table}

\subsection{Simulation 4. Impact of the number of imputations}

A potential difficulty of applying multiple imputation is to make
an appropriate decision on how many imputations are needed. \foreignlanguage{american}{For
example, \citet{Rubin1987} has suggested that five imputations are
sufficient in the case of 50\% missing data for estimating simple
mean. But \citet{Grahametal2007} recommend that many more imputations
than that Rubin recommended should be used. }Although one may always
choose to use a very large number of imputations for mediation analysis
with missing data, this may not be practically possible because of
the amount of computational time involved (In total, K (number of
imputations) x B (number of bootstrap samples) mediation models need
to be estimated). 

In this simulation study, we will briefly investigate the impact of
the number of imputations on the point estimates and standard error
estimates of mediation effects in mediation analysis with missing
data. More specifically, we collect the results from MNAR data analysis
with auxiliary variables with the number of imputations from 10 to
100 with an interval of 10. We focus on how the mediation effect estimates
and the bootstrap standard error estimates change with the number
of imputations. For the purpose of comparison, we calculate the relative
deviances of mediation effect estimates and their standard error estimates
from those estimates with 100 imputations. Those relative deviances
from conditions 10\% missing data and 40\% missing data are plotted
in Figure \ref{fig:imputation}.

Figure \ref{fig:p10} portrays the relative deviances from results
with 10\% missing data. Note that with the number of imputations larger
than 50, the relative deviances of point estimates are all close to
zero and remain unchanged. Thus, 50 imputations seem to be sufficient
for mediation analysis with 10\% missing data. With 40\% missing data,
however, the relative deviances of point estimates do not approach
zero until the number of imputations is larger than 80 as shown in
Figure \ref{fig:p40}. Therefore, the number of imputations required
is related to the amount of missing data.  In our simulation study,
the choice of 100 imputations appears to be enough based on this simulation.

\begin{center}
\begin{figure}[h]
\begin{centering}
\subfloat[10\% missing data]{\begin{centering}
\includegraphics[scale=0.8]{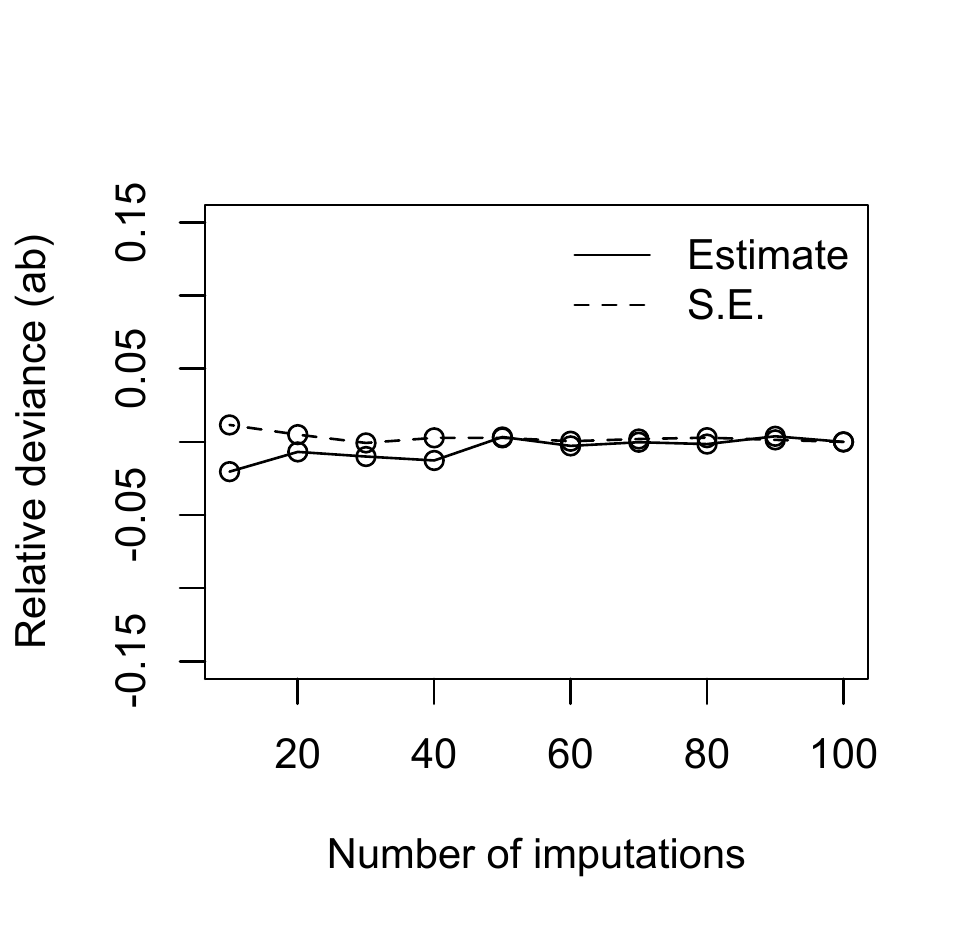}
\par\end{centering}

\centering{}\label{fig:p10}}\subfloat[40\% missing data]{\begin{centering}
\includegraphics[scale=0.8]{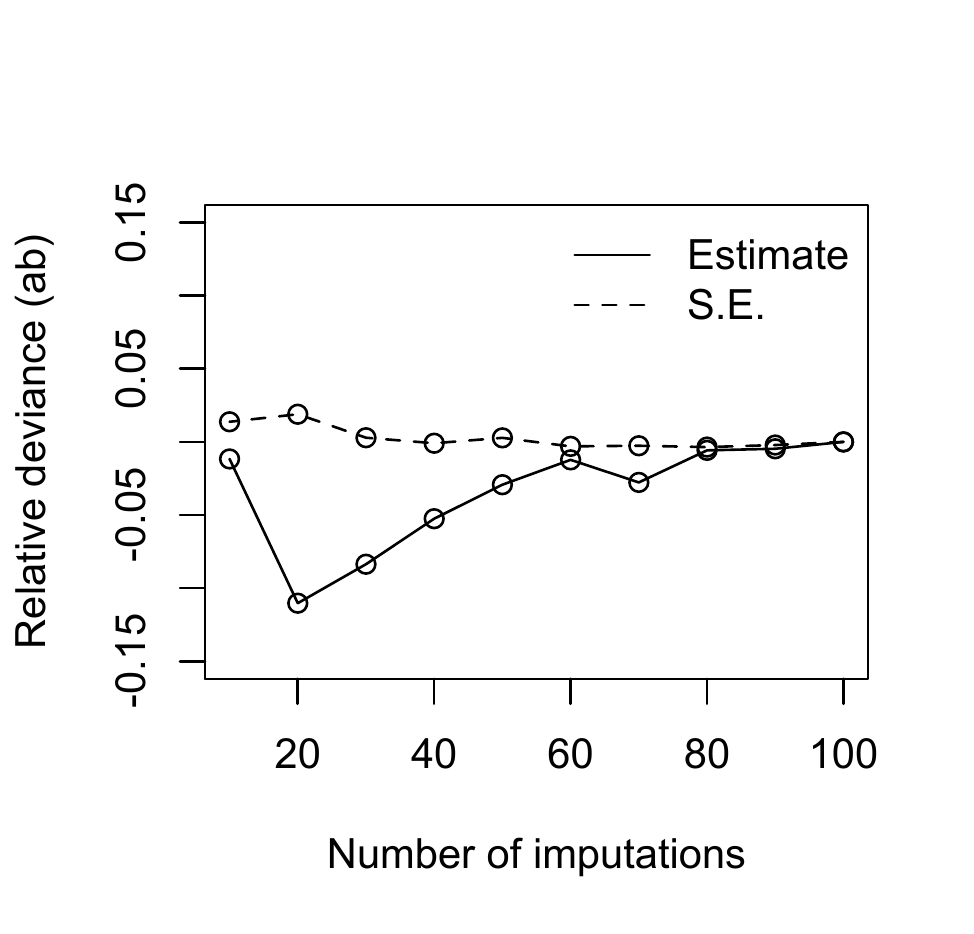}
\par\end{centering}

\centering{}\label{fig:p40}}
\par\end{centering}

\caption{The impact of different numbers of imputations on the accuracy of
point estimates and bootstrap standard error estimates.}
\label{fig:imputation}
\end{figure}

\par\end{center}

\section{An Empirical Example}

In this section, we apply the proposed method to analyze a real data
set to illustrate its application. Research has found that parents'
education levels can influence adolescent mathematics achievement
directly and indirectly. For example, \citet{DavisKean2005} showed
that parents' education levels are related to children's academic
achievement through parents' beliefs and behaviors. To test a similar
hypothesis, we investigate whether home environment is a mediator
in the relation between mothers' education and children's mathematical
achievement . 

Data used in this example are from the National Longitudinal Survey
of Youth, the 1979 cohort \citep[NLSY79,][]{nlsy79}. Data were collected
in 1986 from $N=475$ families on mothers' education level (ME), home
environment (HE), children's mathematical achievement (Math), children's
behavior problem index (BPI), and children's reading recognition and
reading comprehension achievement. For the mediation analysis, mothers'
education is the independent variable, home environment is the mediator,
and children's mathematical achievement is the outcome variable. The
missing data patterns and the sample size of each pattern are presented
in Table \ref{tab:ex:misspat}. In this data set, 417 families have
complete data and 58 families have missing data on at least one of
the two model variables: home environment and children's mathematical
achievement. For the purpose of demonstration, children's behavior
problem index (BPI) and children's reading recognition and reading
comprehension achievement- are used as auxiliary variables in the
data analysis. 

\begin{table}[ht]
\caption{Missing data patterns of the empirical data set.}

\label{tab:ex:misspat} %
\begin{tabular}{lcccc}
\hline 
Pattern  & ME & HE & Math & Sample size \tabularnewline
\hline 
1  & O  & O  & O  & 417\tabularnewline
2  & O  & X  & O  & 36 \tabularnewline
3  & O  & O  & X & 14\tabularnewline
4  & O  & X  & X  & 8\tabularnewline
\hline 
Total  &  &  &  & 475 \tabularnewline
\hline 
\end{tabular}

\emph{Note}. O: observed; X: missing. ME: mother's education level;
HE: home environments; Math: mathematical achievement.
\end{table}

In Table \ref{tab:ex1:res}, the results from empirical data analysis
using the proposed method without and with the auxiliary variables
are presented.%
\footnote{For the empirical data analysis, 1000 bootstraps and 100 imputations
were used.%
} The results reveal that the inclusion of the auxiliary variable only
slightly changed the parameter estimates, standard errors, and the
BC confidence intervals. This indicates that the auxiliary variables
may not be related to the missingness in the mediation model variables.The
results from the analysis with auxiliary variables also show that
home environment partially mediates the relationship between mothers'
education and children's mathematical achievement because both the
indirect effect $ab$ and the direct effect $c'$ are significant. 

\begin{table}[ht]
\caption{Mediation effect of home environment on the relationship between mothers'
education and children's mathematical achievement}

\label{tab:ex1:res} %
\begin{tabular}{l|cccc|cccc}
\hline 
 & \multicolumn{4}{c||}{Without Auxiliary Variable} & \multicolumn{4}{c}{With Auxiliary Variable}\tabularnewline
\hline 
Parameter & Estimate & S.E. & \multicolumn{2}{c||}{95\% BC} & Estimate & S.E. & \multicolumn{2}{c}{95\% BC}\tabularnewline
\hline 
$a$ & 0.035 & 0.049 & 0.018 & 0.162 & 0.036 & 0.049 & 0.018 & 0.163\tabularnewline
$b$ & 0.475 & 0.126 & 0.252 & 0.754 & 0.458 & 0.125 & 0.221 & 0.711\tabularnewline
$c'$ & 0.134 & 0.191 & 0.071 & 0.611 & 0.134 & 0.188 & 0.072 & 0.609\tabularnewline
$ab$ & 0.017 & 0.021 & 0.005 & 0.071 & 0.016 & 0.021 & 0.005 & 0.067\tabularnewline
$i_{Y}$ & 7.953 & 2.047 & 3.530 & 9.825 & 8.045 & 2.025 & 3.778 & 10.006\tabularnewline
$i_{M}$ & 5.330 & 0.556 & 3.949 & 5.641 & 5.327 & 0.558 & 3.945 & 5.646\tabularnewline
$\sigma_{eY}^{2}$ & 4.532 & 0.269 & 4.093 & 5.211 & 4.520 & 0.268 & 4.075 & 5.141\tabularnewline
$\sigma_{eM}^{2}$ & 1.660 & 0.061 & 1.545 & 1.789 & 1.660 & 0.061 & 1.542 & 1.790\tabularnewline
\hline 
\end{tabular}

\emph{Note}. S.E.: bootstrap standard error. BC: bias-corrected confidence
interval.
\end{table}

\section{Discussion}

In this study, we discussed how to conduct mediation analysis with
missing data through multiple imputation and bootstrap. We implemented
the method by using SAS and the program scripts are also provided
and easy to use. Through simulation studies, we demonstrated that
the proposed method performed well for both MCAR and MAR without and
with auxiliary variables. It is also shown that multiple imputation
worked equally well for MNAR if auxiliary variables related to missingness
were included. The analysis of a subset of data from the NLSY79 revealed
that home environment partially mediated the relationship between
mothers' education and children's mathematical achievement.

\subsection{Strength of the proposed method}

The multiple imputation and bootstrap method for mediation analysis
with missing data has several advantages. First, the idea of imputation
and bootstrap is easy to understand. Second, multiple imputation has
been widely implemented in both free and commercial software and thus
can be extended to mediation analysis. %
Third, it is natural and easy to include auxiliary variables in multiple
imputation for analyzing MNAR data. Fourth, multiple imputation does
not assume a specific model for imputing data.

The implementation of multiple imputation and bootstrap in SAS also
has its own advantages. First, only a minimum number of parameters
usually need to be changed to run the SAS program for mediation analysis
with missing data. Second, the SAS program can be easily extended
for more complex mediation analysis by taking advantage of available
SAS procedures. For example, one can also conduct mediation analysis
with moderators through modifying the PROC REG statements. One can
conduct mediation analysis with latent variables through the use of
SAS PROC CALIS. Third, SAS excels in terms of performance in dealing
with large dataset, which is critical for multiple imputation and
bootstrap. For example, for a data set with a sample size 100, to
generate 1000 bootstrap samples and impute each bootstrap sample 100
times, one needs to deal with a data set with 10,000,000 (ten million)
records. Although this seems to be a huge data set, it only took SAS
about 7 minutes to conduct such missing data mediation analysis with
20\% missing data.

\subsection{Assumptions and limitations}

There are several assumptions and limitations of the current study.
First, the study only discusses the mediation model with a single
mediator. The current SAS program is also based on this model. Second,
in applying multiple imputation, we have assumed that all variables
are multivariate normally distributed. However, it is possible that
one or more variables are not normally distributed. Third, the current
mediation model only focuses on the cross-sectional data analysis.
Some researchers have suggested that the time variable should be considered
in mediation analysis \citep[e.g., ][]{Cole2003,MacKinnon2008,Wang2009}.
Fourth, in dealing with MNAR data, we assume that useful auxiliary
variables that can explain missingness in the mediation model variables
are available . However, sometimes the auxiliary variables may not
be available. 

In summary, a method using multiple imputation and bootstrap for mediation
analysis with missing data is introduced and the program of implementing
this method is developed in SAS. Simulation results show that the
method works well in dealing with missing data for mediation analysis
under different missing mechanisms. We hope this program can promote
the use of advanced techniques in dealing with missing data for mediation
analysis in the future. 

\bibliographystyle{apalike}

\appendix

\section{SAS codes for MI and bootstrap}

\linespread{1}

\begin{lstlisting}[basicstyle={\footnotesize\ttfamily},breaklines=true,language=SAS,numbers=left,numberstyle={\footnotesize}]
/*** Setup the global parameters ***/
/*The parameters below should be changed accordingly*/
%LET filename="c:\mnarmediation\dataname.txt"; * data file directory and name;
%LET varname=x m y a1 a2;  *specify variable names in the data file. Please use x for the input variable, m for the mediation variable, and y for the output variable. a1 and a2 are two auxiliary variables in the data file. You can use any names except for x, m, and y for naming the auxiliary variables;
%LET missing=99999; *specify the missing data value;
%LET nimpute = 100; *define the number of imputations K;
%LET nboot = 1000;  *define the number of bootstraps B;
%LET alpha = 0.95;  *define the confidence level;
%LET seed = 2010;   *random number seed;
/*** End of setup of global parameters ***/


/*In general, there is no need to change the codes below*/
/*Read data into sas*/
DATA dset;
  INFILE &filename;
  INPUT &varname;
  ARRAY nvarlist &varname;
  DO OVER nvarlist;
    IF nvarlist = &missing THEN nvarlist = .;  
  END;
RUN;

/*Use multiple imputation to obtain point estimates of the model parameters  based on the original data set*/
/*Imputing the original data set multipe times*/
PROC MI DATA=dset SEED=&seed NIMPUTE=&nimpute OUT=imputed NOPRINT;
  VAR &varname;
RUN; QUIT;
/*Estimating model parameters for each imputed data set*/
PROC REG DATA=imputed OUTEST= est NOPRINT;
  MODEL y = x m;
  MODEL m = x;
  BY _Imputation_;
RUN; QUIT;

/*Collecting results from mutiple imputations*/
DATA temp;
  SET est;
  id =INT((_N_-.1)/2)+1;
  modelnum = MOD(_N_+1, 2)+1;
RUN;

DATA temp1;
  SET temp;
  ARRAY int[2] iY iM;
  ARRAY xpar[2] c a;
  ARRAY mpar[2] b tmp1;
  ARRAY sigma[2] sy sm;
  RETAIN  a b c iY iM sy sm;
	BY id;
	IF FIRST.id THEN DO I = 1 to 2;
		int[I] = .;
		xpar[I] = .;
		mpar[I]=.;
		sigma[I]=.;
	END;
	int[modelnum] = intercept;
	xpar[modelnum] = x;
	mpar[modelnum] = m;
	sigma[modelnum] = _RMSE_;
	IF LAST.id THEN OUTPUT;
	KEEP  _imputation_ a b c iY iM sy sm;
RUN;
/*Calcuating mediation effects*/
DATA temp2;
  SET temp1;
  ab=a*b;
RUN;

/*Saving the point estimates of model parameters and mediation effect from multiple imputation into a data set named 'pointest'*/
PROC MEANS DATA=temp2 NOPRINT;
  VAR a b c ab iY iM sy sm;
  OUTPUT OUT=pointest MEAN(a b c ab iY iM sy sm)=a b c ab iY iM sy sm;
RUN;

/*** Bootstraping data to obtain standard errors and confidence intervals ***/
DATA bootsamp;
  DO sampnum = 1 to &nboot;
     DO i = 1 TO nobs;
        ran = ROUND(RANUNI(&seed) * nobs);
        SET dset
        nobs = nobs
        point = ran;
        OUTPUT;
     END;
  END;
  STOP;
RUN; QUIT;

/*** Imputing K data sets for each bootstrap sample ***/
PROC MI DATA=bootsamp SEED=&seed NIMPUTE=&nimpute OUT=imputed NOPRINT;
  EM MAXITER = 500;
  VAR &varname;
  BY sampnum;
RUN; QUIT;

/*Estimate model parameters for each imputed data set (in total, there are B*K imputed data sets.)*/
PROC REG DATA=imputed OUTEST= est NOPRINT;
  MODEL y = x m;
  MODEL m = x;
  BY sampnum _Imputation_;
RUN; QUIT;

/*Collecting results from different imputed data sets*/
DATA temp;
  SET est;
  id =INT((_N_-.1)/2)+1;
  modelnum = MOD(_N_+1, 2)+1;
RUN;

DATA temp1;
  SET temp;
  ARRAY int[2] iY iM;
  ARRAY xpar[2] c a;
  ARRAY mpar[2] b tmp1;
  ARRAY sigma[2] sy sm;
  RETAIN  a b c iY iM sy sm;
	BY id;
	IF FIRST.id THEN DO I = 1 to 2;
		int[I] = .;
		xpar[I] = .;
		mpar[I]=.;
		sigma[I]=.;
	END;
	int[modelnum] = intercept;
	xpar[modelnum] = x;
	mpar[modelnum] = m;
	sigma[modelnum] = _RMSE_;
	IF LAST.id THEN OUTPUT;
	KEEP  sampnum _imputation_ a b c iY iM sy sm;
RUN;

DATA temp2;
  SET temp1;
  ab=a*b;
RUN;

/*Compute point estimates of model parameters and mediation effect for each bootstrap sample and the results are saved in the data file named 'bootest'. */
PROC MEANS DATA=temp2 NOPRINT;
  BY sampnum;
  VAR a b c ab iY iM sy sm;
  OUTPUT OUT=bootest MEAN(a b c ab iY iM sy sm)=a b c ab iY iM sy sm;
RUN;

/*** Calculate the BC intervals based on the point estimates from different bootstrap samples and produce a table containing the points estimates, standard errors, confidence intervals in the output window.***/
PROC IML;
  START main;
  USE pointst;
  READ ALL INTO Y;
  USE bootest;
  READ ALL INTO X;

  n=NROW(X);
  m=NCOL(X);

  bc_lo=J(1,m-3,0);
  bc_up=J(1,m-3,0);
  se=J(1,m-3,0);

  alphas=1-(1-&alpha)/2;
  zcrit = PROBIT(alphas);
  
  DO j=1 TO m-3;
	se[j]=SQRT((SSQ(X[,j+3]) -(SUM(X[,j+3]))**2/n)/(n-1)); 
	number=0;
	DO i=1 TO n;
		IF X[i,j+3]<Y[j+2] THEN number=number+1;
	END;
	p=number/n;
	z0hat=PROBIT(p);
        
	q1=z0hat+(z0hat-zcrit);
	q2=z0hat+(z0hat+zcrit);
	alpha1=PROBNORM(q1);
	alpha2=PROBNORM(q2);

	vec=X[,j+3];
	CALL SORT(vec,{1});

	low=int(alpha1*(n+1));
	up=int(alpha2*(n+1));
	IF low<1 THEN low=1;
	IF up>n THEN up=n;
	bc_lo[j]=vec[low];
	bc_up[j]=vec[up];
  END;

  result=Y[3:10]||se`||(bc_lo`)||(bc_up`);
  MATTRIB result ROWNAME=({a, b, c, ab, iy, im, sy, sm})
                 COLNAME=({estiamtes se CI_lo CI_up})
                 LABEL='MEDIATION ANALYSIS RESULTS' FORMAT=f10.5;
  PRINT result;
  FINISH;
  RUN main;
QUIT;
\end{lstlisting}

\end{document}